# Comment on “Negative Refractive Index in Artificial Metamaterials” [A. N. Grigorenko, Opt. Lett., 31, 2483 (2006)]

**Alexander V. Kildishev°, Vladimir P. Drachev°, Uday K. Chettiar°, Douglas Werner°°, Do-Hoon Kwon°°, and Vladimir M. Shalaev°**

°Birck Nanotechnology Center, Purdue University, West Lafayette, IN 47907, USA

°°Department of Electrical Engineering, Pennsylvania State University, University Park, PA 16804, USA

A key optical parameter characterizing the existence of negative refraction in a thin layer of a composite material is the effective refractive index of an equivalent, homogenized layer with the same physical thickness as the initial inhomogeneous composite. Measuring the complex transmission and reflection coefficients is one of the most rigorous ways to obtain this parameter. We dispute Grigorenko’s statement (Optics Letters 31, 2483 (2006)) that measuring only the reflection intensity spectrum is sufficient for determining the effective refractive index. We discuss fundamental drawbacks of Grigorenko’s technique of using a best-fit approximation with an *a priori* prescribed dispersive behavior for a given metamaterial and an ‘effective optical thickness’ that is smaller than the actual thickness of the sample. Our simulations do not confirm the Grigorenko paper conclusions regarding the negative refractive index and the negative permeability of the nanopillar sample in the visible spectral range.

An alternative way to obtain the effective parameters ($\mu$ and $\varepsilon$, and thence $n$) of a thin layer of a sub-wavelength periodic metamaterial has been reported recently by Grigorenko.[1,2] The core of the reported method is built on a best-fit procedure of a given reflection spectra with *a priori* prescribed dispersion relations for $\varepsilon(\lambda)$ and $\mu(\lambda)$. The method uses six fitting constants and an 'effective optical thickness' of the sample, which is smaller than the actual, physical thickness of the sample. The Grigorenko method (GM) seems to be much simpler than other methods in validating thin metal-dielectric composites as possible prototypes for a negative index metamaterial (NIM). We note that normally both the transmission and reflection spectra are involved in measurements, assisted by direct or indirect measurements of the phases acquired in transmission and reflection.[3,4] For example, the first stage of optical characterization can be provided by the spectral analysis of the complex transmission and reflection coefficients, $t$ and $r$,[5-7] further followed by retrieval of the effective refractive index, $n = n' + \imath n''$ and impedance, $\eta = \eta' + \imath \eta''$.[4,6-8] This technique gives unambiguous results for the prime parameters ($n$ and $\eta$) and the associated effective values of permeability, $\mu = n\eta$, and permittivity, $\varepsilon = n/\eta$.

In Ref. [2], the author claims that the best sample of Au nanopillars covered with a thin glycerine layer shows a negative index of refraction $n = -0.7$ at wavelengths corresponding to green light and with a quality ratio $-n'/n'' = 0.4$. Interestingly, the claim in Ref. [2] contradicts the original message in Ref. [1] regarding the same sample: "Although our structures exhibited both negative $\mu$ and negative $\varepsilon$ within the same range of $\lambda$ (for example, $\varepsilon' \approx -0.7$ and $\mu' \approx -0.3$ at the green resonance …, $\mu$ had a rather large imaginary component ($\mu'' \equiv \mathrm{Im}(\mu) \approx 1i$ at the resonance), which so far has not allowed the observation of negative refraction." Generally, it really does not matter whether $\mu''$ is large or not; since $\mu'|\varepsilon|+\varepsilon'|\mu|<0$ is always sufficient for $n'<0$, $n'$ would be negative provided that $\varepsilon'\approx -0.7$ and $\mu'\approx -0.3$.

Our Comment was inspired by these controversial statements from the two papers.[1,2] It is worth pointing out that in contrast to their claims; only a positive refractive index was obtained in our validation tests for the same geometry. Additionally, neither negative magnetic response nor negative electric response at green light was obtained in the simulations using three different simulation tools: (1) a combination of periodic finite element method - boundary integral (PFEBI) solver,[9] (2) a commercial finite element software (COMSOL), and (3) a parallel 3D FDTD solver. A unit cell of the array, also used in our validation, is shown in Fig. 1a; meshed unit cell for both of our frequency domain solvers are shown in Fig. 1ab. The rectangular uniform grid of the FDTD solver is similar to the mesh used for PFEBI (Fig. 1b).

In both works[1,2] the discussion is focused on the characterization of a single layer of coupled gold 'nanopillars' arranged in a bi-periodic array with sub-wavelength periodicity. Both papers[1,2] refer to a full-wave numerical simulation using a commercial finite element solver (FEMLAB, now COMSOL software). The authors use a Drude model with the parameters for gold shown in Ref. [1, p. 337]. We note that the Drude model is incapable of adequately describing the behavior of gold in the visible wavelength range. For example, Fig. 2 compares the results obtained using the Drude formula with the parameters from Ref. [5,10] versus the experimental data.[11,12] Although the match at the wavelength range above the visible is quite good, the area corresponding to green light differs substantially. We therefore comment that in the visible range, the results of numerical simulations shown in Ref. [1] are of little assistance to the problem.

Now we consider the alternative approach first demonstrated in Ref. [1], and then discussed in some detail in Ref. [2]. In essence, the approach follows a four-step recipe:

Step 1: Determine a reflectance spectrum ($R(\lambda)$) of a given thin sample from either measurements or simulations;

Step 2: Prescribe fixed dispersive relationships for both $\mu(\lambda)$ and $\varepsilon(\lambda)$ using the following formulae [2]:

$$\begin{aligned}\mu(\lambda) &= 1 + F_m\lambda_m^2 \big/ \left(\lambda^2 - \lambda_m^2 - \iota\lambda\Delta\lambda_m\right),\\ \varepsilon(\lambda) &= 1 + F_e\lambda^2 \big/ \left(\lambda^2 - \lambda_e^2 - \iota\lambda\Delta\lambda_e\right);,\end{aligned} \qquad (1)$$

Step 3: Match the reflection spectra using six fitting parameters ($F_m$, $\lambda_m$, $\Delta\lambda_m$, $F_e$, $\lambda_e$, and $\Delta\lambda_e$) and an 'effective optical thickness.' Note that the effective thickness is defined separately using either ellipsometry or taken as a 'mass thickness,' i.e. is much smaller than the actual physical thickness of the layer of nanopillars. (The physical thickness of the layer is denoted as $h$ in Fig. 1a.)

Step 4: Finally, calculate the effective refractive index using the following formula Ref. [2, Eq. (2)] with the fitting parameters obtained at Step 3.

$$n(\lambda) = \sqrt{\varepsilon(\lambda)}\sqrt{\mu(\lambda)}sign\left(\sqrt{\varepsilon(\lambda)}\big/\sqrt{\mu(\lambda)}\right), \qquad (2)$$

Below we compare this GM approach with our simulations. The description of the approach is incomplete in Refs. [1,2]. In order to verify that we understood the approach correctly, we compared the approximated spectrum of $R$ calculated using the original fitting parameters (listed in the caption of Ref. [2, Fig. 2b]) with the same spectrum obtained from measurements also shown in Ref. [2, Fig. 2b]. Indeed, the fitting parameters selected by the author in Ref. [2] provide an adequate match with the experimental curve as shown in Fig. 3a. Panels b, c, and d of Fig. 3 also show the results of the formal calculations using formulas (1) and (2). However the results of the GM approach contain an internal inconsistency (to be shown presently) and differ strongly from our electromagnetic simulations.

To validate the results, a hybrid technique using a combination of periodic finite element method - boundary integral (PFEBI) technique has been utilized to obtain the spectra of $t$ and $r$; then, a standard method[3,4,7,8] has been used first for obtaining $n$ and $\eta$, and then, $\mu$ and $\varepsilon$. The results obtained from PFEBI are also consistent with our calculations using a commercial finite element software (COMSOL Multiphysics) and our 3D FDTD solver. Since the specific effects (a magnetic response and even a negative refractive index) are claimed by Grigorenko at green light, the spectra shown in Fig. 3 are centered around the point of interest. The following geometrical parameters $a = 400$ nm, $s = 140$ nm, $h = 90$ nm for our simulations are taken directly from Ref. [1,2], while $d_1 = 93$ nm and $d_2 = 127$ nm are obtained using the averaging of selected measurements of zoomed micrographs and our estimates of possible slopes associated with electron beam lithography. The simulation takes into account the glass substrate; experimental optical constants of bulk gold are used for the material of the nanopillars.[11] While the reflectance spectra for the measured, best-fit, and simulated results agree relatively well (Fig. 3a), further comparison indicates strong difference between the values of the effective parameters (panels b, c, and d in Fig 3).

The major ambiguity in Grigorenko's development is that neither the phase changes of the reflected light nor the transmission spectra (and phase changes in transmission) are considered in his homogenization procedure.[1,2] Thus, the reflectance spectra alone is believed to carry sufficient information for the restoration of the complex values of $\mu$ and $\varepsilon$. Further, an 'effective optical thickness' is used throughout both papers instead of the physical thickness normally taken in standard homogenization techniques. Since the effective optical thickness is much smaller than the actual, physical thickness (e.g., 12 nm vs. 90 nm, see Supplementary Information, p. 3 in Ref. [1]), any resonant effects obtained using this effective thickness are therefore much stronger and leave any sample almost no chance of deviating from the behavior most sought after by the author(s) in Refs. [1,2].

Although the formulae (1) are shown in Refs. [1,2], their possible artifacts are not discussed anywhere in the two papers. Specifically, the knowledge of the complex values of $\mu$ and $\varepsilon$ obtained from a given spectrum of $R$ suggests that an unknown spectrum of $T$ also could be restored. Fig. 4 shows an example of the 'restored' transmittance obtained using the original fitting parameters ($F_m$, $\lambda_m$, $\Delta\lambda_m$, $F_e$, $\lambda_e$, and $\Delta\lambda_e$) and the 'effective thickness' compared with the result obtained for the same sample using our simulations and the standard restoration technique. The restored $T$ (open squares in Fig. 4) exhibits unusual behavior, since any magnetic resonance due to localized plasmonic effects is expected to accompany relatively sharp enhanced absorbance at the resonance position.[3,13] Regrettably, this is not the case for the restored values of $T$. This discrepancy could have been easily identified provided that the transmittance had been measured in the experiments.

Unfortunately, our numerical simulations show that the retrieval of the effective permeability ($\mu'$) and permittivity ($\varepsilon'$) for the glycerine-covered sample in Ref. [1] is also inconclusive, and the claim for a negative refractive index in the visible made in Ref. [2] is not well grounded. The disagreement is clearly shown in Fig. 5a-b.

Upon realizing that the simulations performed for this glycerine-covered sample indicated neither a negative index of refraction nor a negative effective permeability, we made an additional straightforward analysis of Ref. [1] that revealed further albeit less significant mismatched details. For instance, suppose we follow the ideology of Refs. [1,2] and use the fitting parameters ($F_m$, $\lambda_m$, $\Delta\lambda_m$, $F_e$, $\lambda_e$, and $\Delta\lambda_e$) from Ref. [1, Fig. 4] for the glycerine-covered sample since it has been claimed to produce negative refraction. Interestingly, even if we follow the recipe of Refs. [1,2] and get similar results for the refractive index, permittivity and permeability (shown in Figs. 6a-c), we see a substantial discrepancy in the best-fit approximation of the reflection spectra. While, as indicated in Fig. 6d, the ends of the experimental curve are matched quite well, the major mismatch occurs at the region of main interest, i.e. around the 'green light resonance.' Additionally, 'ideal' resonant curves of Figs. 6a-c are not consistent with the moderate behavior of the similar effective parameters shown in Fig. 5b.

To further illustrate the misleading 'reflectance best-fit concept' of Refs. [1, 2], we have made a simple experiment using a standard absorbing filter with a reflectance spectrum similar to the nanopillars sample. First, a reflection spectra has been measured and matched with formulas (1) assuming an 'effective optical thickness' of 90 nm (supposing all absorbing centers are compacted in a layer of the effective thickness). The matched constants were used then for obtaining the 'effective permeability' and 'effective permittivity'; and then, an 'effective refractive index' was obtained using the same matched constants. Finally, using the same equivalent optical thickness with the complex values of $\mu$ and $\varepsilon$ retrieved from the reflectance spectrum, we restored the transmission spectrum. The test results are shown in Fig. 7a-d. Although the reflection spectrum is matched to the experiment quite well (Fig. 7a), the transmission spectrum is dramatically different from the measurement (Fig. 7b). Amazingly enough, according to the Grigorenko procedure the standard glass filter also shows a negative refractive index and a negative magnetic response, as shown in Fig. 7cd.

It is not our intention to discourage against feasible applications of best-fitting approaches to adequate retrieval of the effective optical constants. For example, a successful optimization approach for retrieving the effective material parameters of a general bianisotropic layer is shown in Ref. [14]. As expected, that approach is using the complete information about both the reflected and transmitted light, i.e. it employs both the magnitudes and phase changes gained in reflection and transmission.

In summary, we show that it is doubtful that a sample with the geometry and materials proposed in Ref. [1,2] and schematically depicted in Fig. 1 is capable of demonstrating either a negative effective permeability ($\mu'$) or a negative effective refractive index. It also seems that even a negative effective permittivity ($\varepsilon'$) at green light ($520-570\ \mathrm{nm}$) is questionable for this structure either with or without a layer of glycerine.

## References (with titles)

## Figure captions

1. Fig. 1. (a) A unit cell of the original design of a bi-periodic array of gold nanoparticles (‘nano-pillars’) placed on top of a thick glass substrate as described in Refs. [1,2]. In Ref. [2] the author claims that the same sample covered with a thin layer of glycerine shows a negative index of refraction. (b) and (c) Meshed unit cells of PFEBI and FEM solvers respectively.
2. Fig. 2. The complex part of permittivity calculated using the parameters of Ref. [5] and the experimental data Refs. [9,12].
3. Fig. 3. (a) Reflectance spectra obtained for the sample of Fig. 1 using a best-fitting technique Ref. [2] vs. the experiment also shown in Ref. [2]. The best-fit results (□) and our simulations (○) match well with the experimental results (●). (b) - (d) Effective parameters ($n'$, $n''$, $\mu'$, $\mu''$, $\varepsilon'$, $\varepsilon''$) obtained from the best-fit (–, --) vs. the same parameters obtained from our numerical simulations (-□-, -Δ-).
4. Fig. 4. Restored transmittance using best-fit results (□) and our simulations (○).
5. Fig. 5. (a) Transmittance and reflectance spectra calculated for the glycerine covered sample using PFEBI. The simulated reflectance spectrum is compared to the experiment of Ref. [1]. (b) The real parts of permittivity, permeability and refractive index calculated from the complex reflection and transmission coefficients for the same sample.
6. Fig. 6. (a)-(c) Effective parameters calculated use the fitting constants from Ref. [1, Fig. 4] for the glycerine-covered sample. (d) Reflectance spectrum obtained from the experimental data of Ref. [1] and calculated using the same best-fit constants.

7. Fig. 7. (a) Reflectance spectra obtained for a standard absorbing glass filter using the best-fit technique using an effective optical thickness and our experimental data. (b) The experimental transmittance spectrum and the spectrum restored using the same effective thickness and fitting constants obtained from the reflectance spectrum. (c) - (d). Effective parameters ($n'$, $n''$, $\mu'$, $\mu''$) obtained from the best-fit approach.

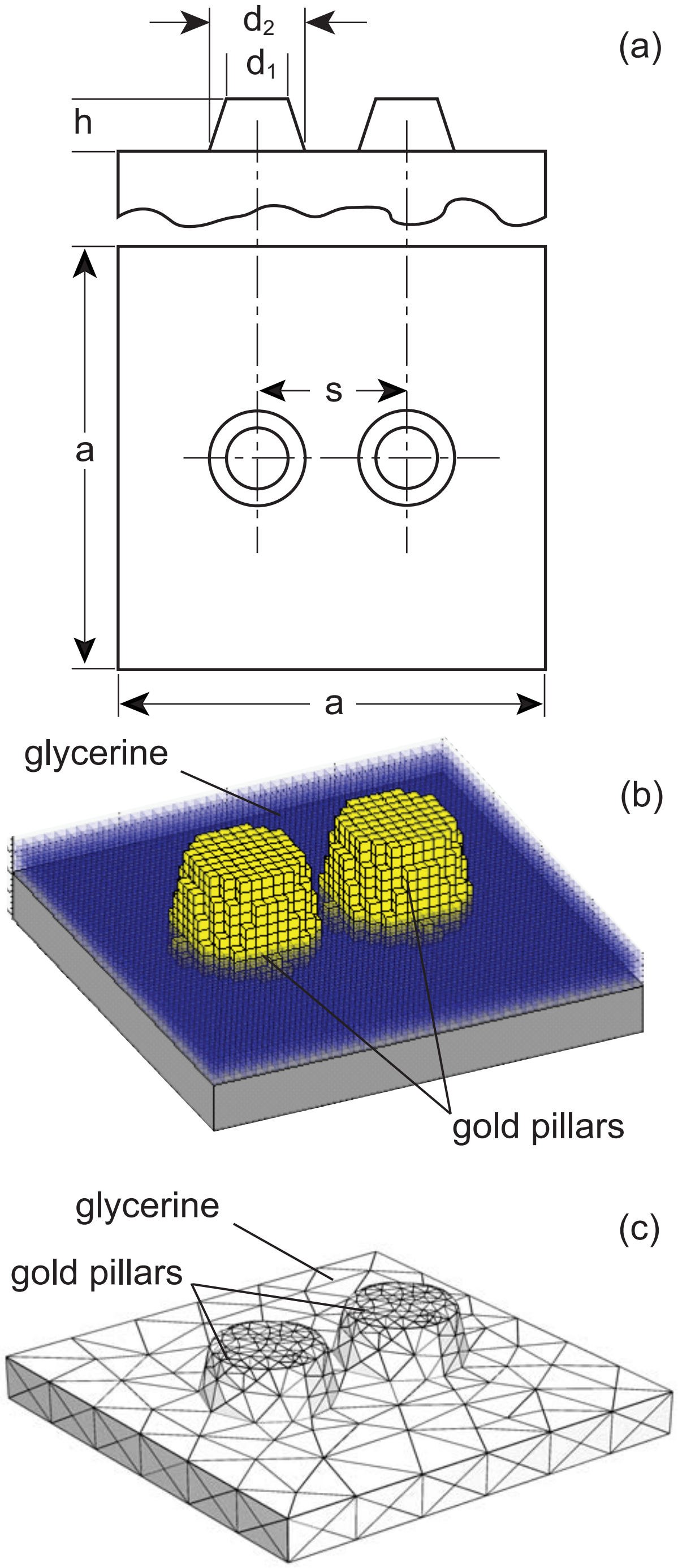

**Fig. 1.** (a) A unit cell of the original design of a bi-periodic array of gold nanoparticles (‘nano-pillars’) placed on top of a thick glass substrate as described in Refs. [1, 2]. In Ref. [2] the author claims that the same sample covered with a thin layer of glycerine shows a negative index of refraction. (b) and (c) Meshed unit cells of PFEBI and FEM solvers respectively.

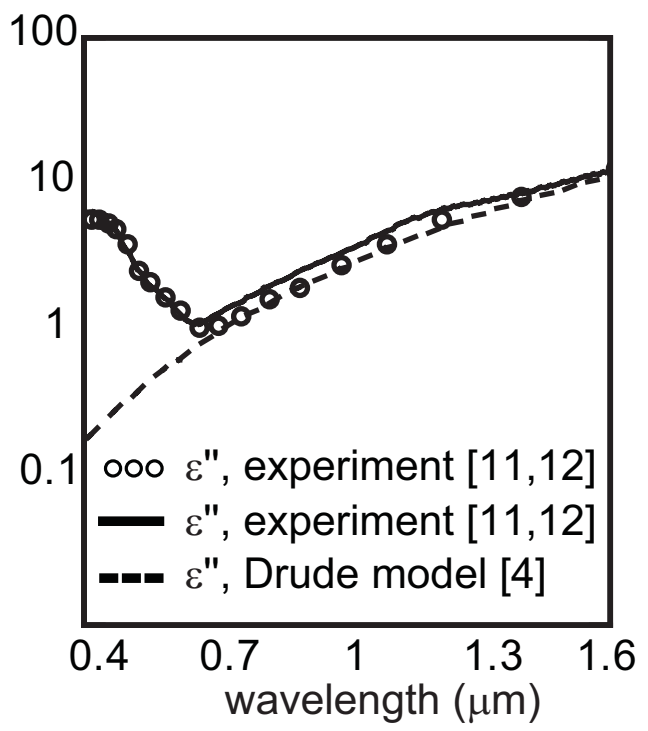

**Fig. 2.** The complex part of permittivity calculated using the parameters of Ref. [5] and the experimental data Refs. [11,12].

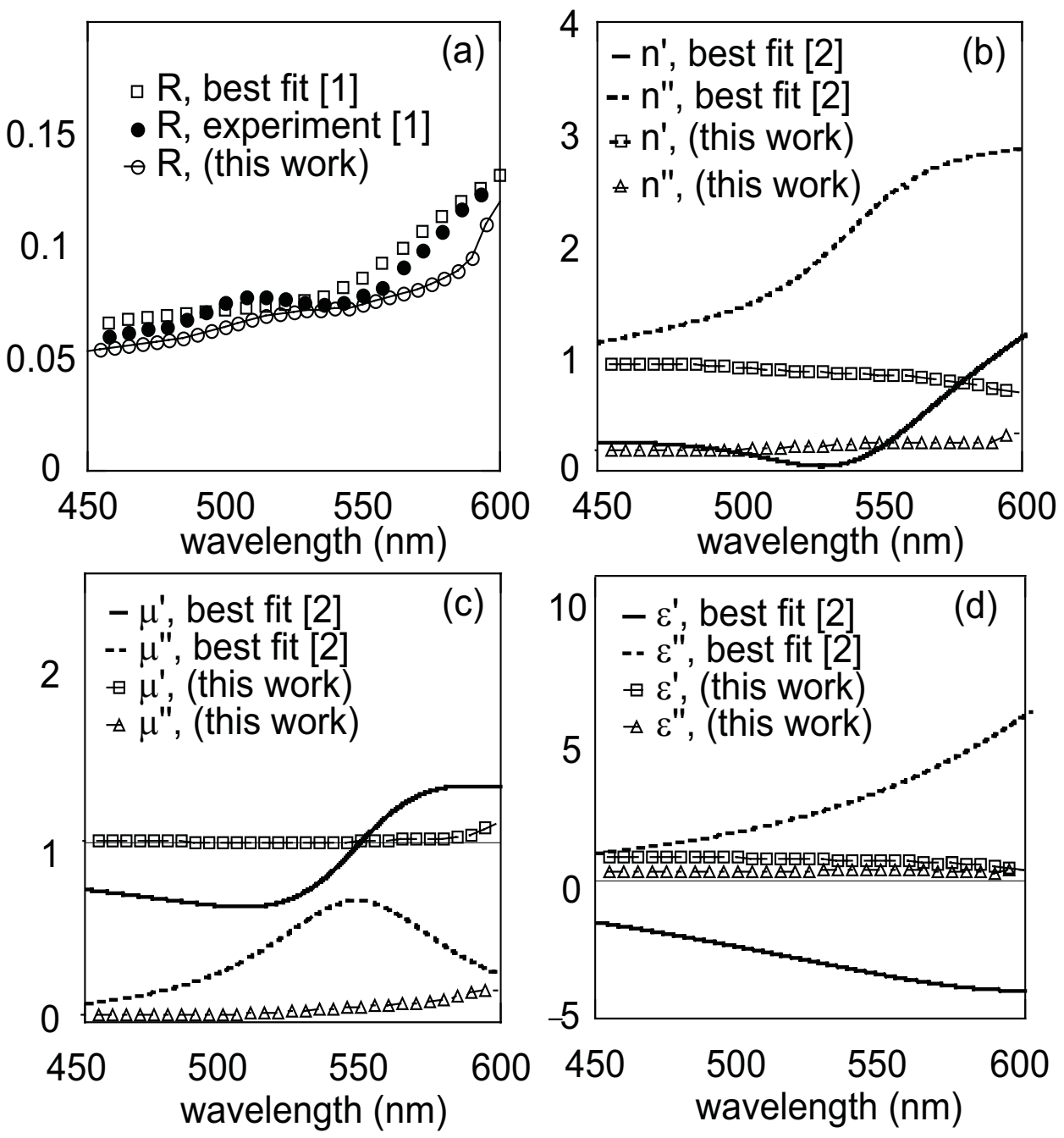

**Fig. 3.** (a) Reflectance spectra obtained for the sample of Fig. 1 using a best-fitting technique Ref. [2] vs. the experiment also shown in Ref. [2]. The best-fit results (□) and our simulations (○) match well with the experimental results (●). (b) - (d) Effective parameters ($n'$, $n''$, $\mu'$, $\mu''$, $\varepsilon'$, $\varepsilon''$) obtained from the best-fit (—, --) vs. the same parameters obtained from our numerical simulations (-□-, -Δ-).

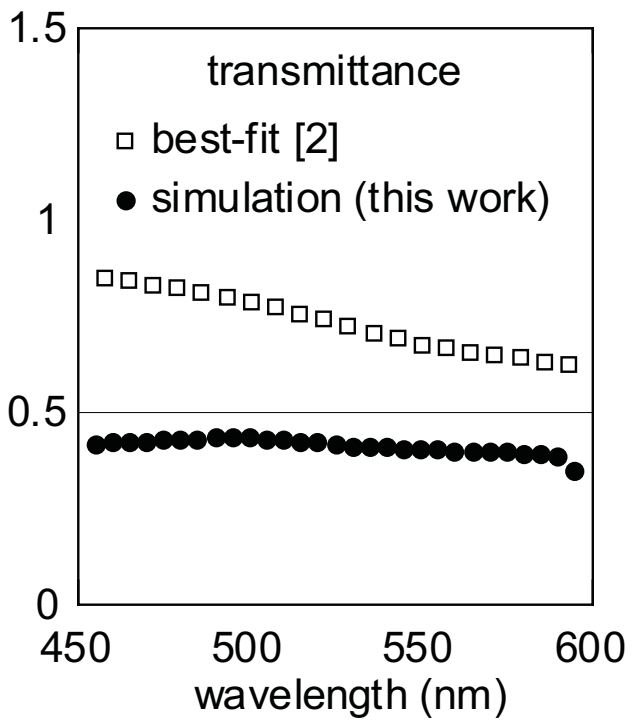

**Fig. 4.** Restored transmittance using best-fit results (□) and our simulations (○).

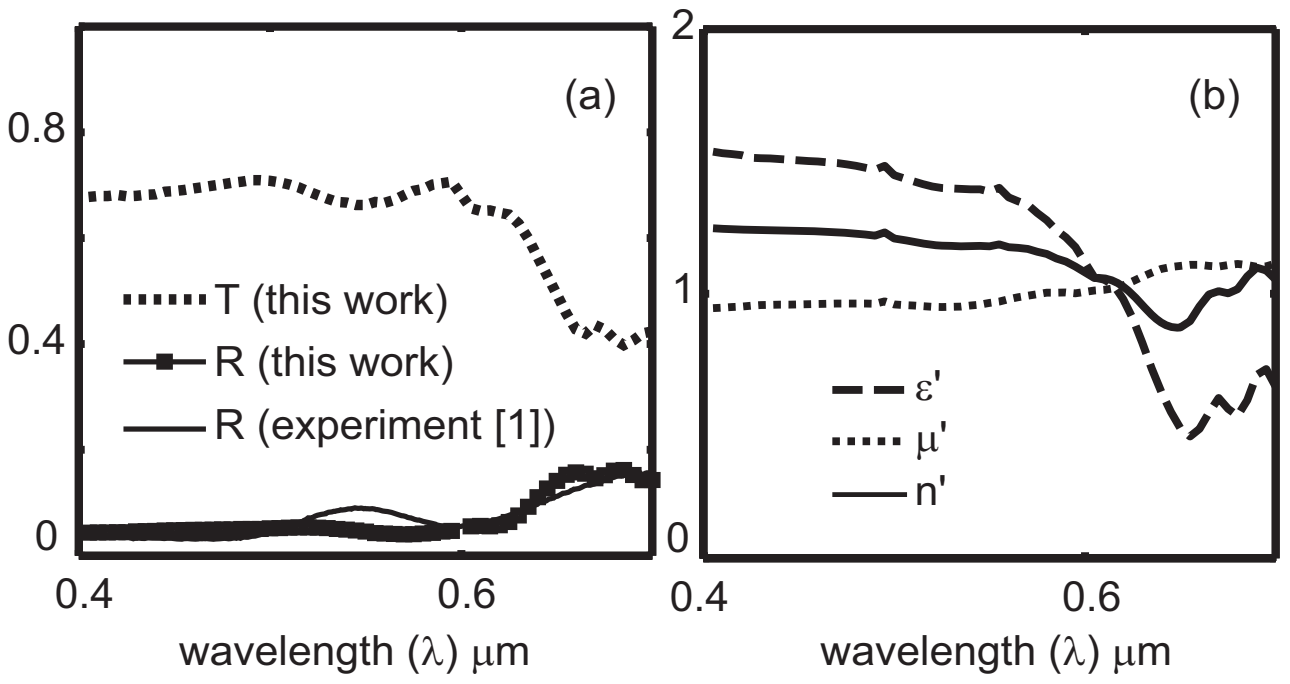

**Fig. 5.** (a) Transmittance and reflectance spectra calculated for the glycerine covered sample using PFEBI. The simulated reflectance spectrum is compared to the experiment of Ref. [1]. (b) The real parts of permittivity, permeability and refractive index calculated from the complex reflection and transmission coefficients for the same sample.

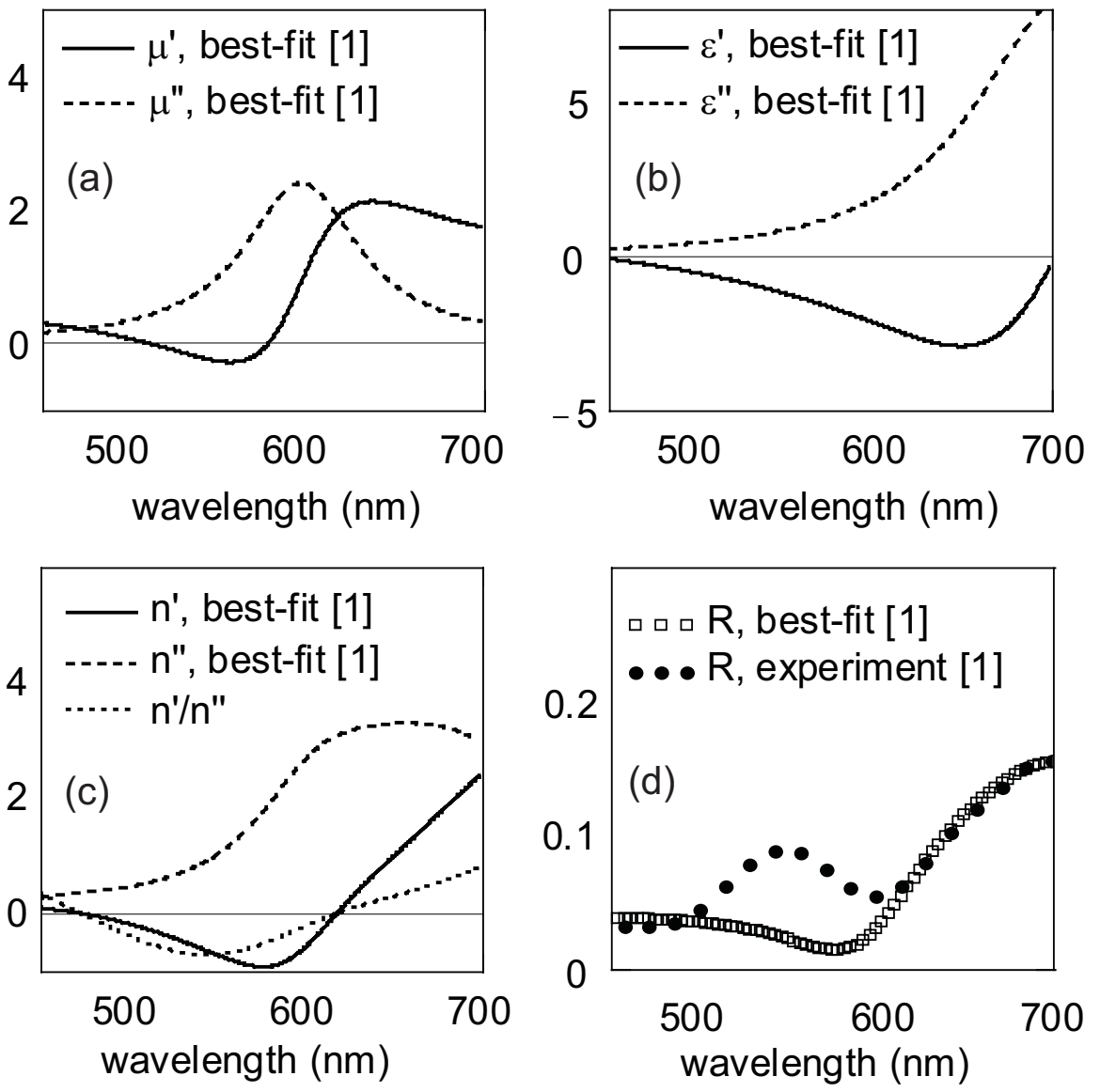

**Fig. 6.** (a)-(c) Effective parameters calculated use the fitting constants from Ref. [1, Fig. 4] for the glycerine-covered sample. (d) Reflectance spectrum obtained from the experimental data of Ref. [1] and calculated using the same best-fit constants.

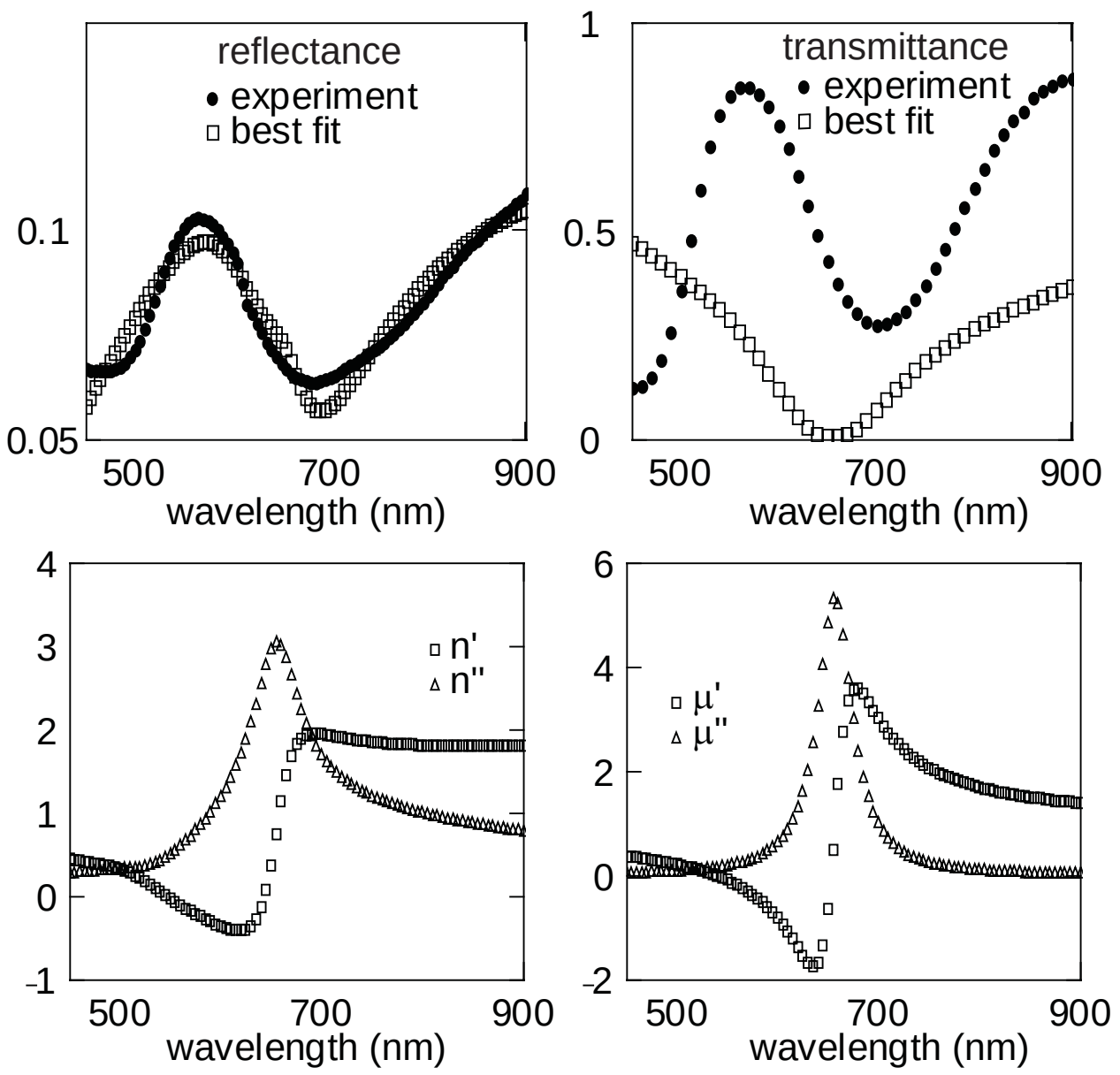

**Fig. 7.** (a) Reflectance spectra obtained for a standard absorbing glass filter using the best-fit technique using an effective optical thickness and our experimental data. (b) The experimental transmittance spectrum and the spectrum restored using the same effective thickness and fitting constants obtained from the reflectance spectrum. (c) - (d). Effective parameters ($n'$, $n''$, $\mu'$, $\mu''$) obtained from the best-fit approach.